\pdfoutput=1

\PassOptionsToPackage{table,x11names,dvipsnames}{xcolor}

\documentclass[11pt]{article}

\usepackage[preprint]{acl}

\usepackage{times}
\usepackage{latexsym}
\usepackage{amsmath}

\usepackage[T1]{fontenc}

\usepackage[utf8]{inputenc}

\usepackage{microtype}

\usepackage{inconsolata}

\usepackage{graphicx}
\usepackage{tikz}
\usepackage{booktabs}

\newcommand{\ray}[1]{
}
\newcommand{\MToolName}{SemAgent-Multi\space}
\newcommand{\ToolName}{SemAgent\space}
\newcommand*\circled[1]{\tikz[baseline=(char.base)]{
            \node[shape=circle,draw,inner sep=0.75pt] (char) {#1};}}
\newcommand*\blackcircled[1]{\tikz[baseline=(char.base)]{
    \node[shape=circle, fill=black, text=white, inner sep=0.75pt, draw=black] (char) {#1};}}

\definecolor{lightmauve}{rgb}{0.86, 0.82, 1.0}

\usepackage{makecell} 
\usepackage{enumitem} 

\usepackage{circledtext}
\usepackage{newtxtext}
\usepackage{tikz}
\usepackage{libertine}

\usepackage{url}

\title{\ToolName{}: A Semantics Aware Program Repair Agent}

\author{  $\text{Anvith Pabba}^{1}$ \quad  $\text{Alex Mathai}^{1}$ \quad $\text{Anindya Chakraborty}^{2}$ \quad 
 $\text{Baishakhi Ray}^{1}$ \\
  ${}^1 \text{Columbia University}$ \\
  $\textsf{\{ap4450, am6215\}@columbia.edu} \quad {}^2 \textsf{\{anindyaju99\}@gmail.com} \quad \textsf{\{rayb\}@cs.columbia.edu}$ \\
}

\usepackage[parfill]{parskip}
\usepackage[most]{tcolorbox}
\usepackage{lipsum}
\usepackage{tcolorbox}

\begin{document}
\maketitle
\begin{abstract}
  Large Language Models (LLMs) have shown impressive capabilities in downstream software engineering tasks such as Automated Program Repair (APR). In particular, there has been a lot of research on repository-level issue-resolution benchmarks such as SWE-Bench \cite{jimenez2024swebenchlanguagemodelsresolve}. Although there has been significant progress on this topic, we notice that in the process of solving such issues, existing agentic systems tend to hyper-localize on immediately suspicious lines of code and fix them in isolation, without a deeper understanding of the issue semantics, code semantics, or execution semantics. Consequently, many existing systems generate patches that overfit to the user issue, even when a more general fix is preferable. 
  To address this limitation, we introduce \ToolName{}, a novel workflow-based procedure that leverages issue, code, and execution semantics to generate patches that are complete -- identifying and fixing all lines relevant to the issue. 
  We achieve this through a novel pipeline that (a) leverages execution semantics to retrieve relevant context, (b) comprehends issue-semantics via generalized abstraction, (c) isolates code-semantics within the context of this abstraction, and (d) leverages this understanding in a two-stage architecture: a repair stage that proposes fine-grained fixes, followed by a reviewer stage that filters relevant fixes based on the inferred issue-semantics. Our evaluations show that our methodology achieves a solve rate of $44.66\%$ on the SWEBench-Lite benchmark beating all other workflow-based approaches, and an absolute improvement of $7.66\%$ compared to our baseline, which lacks such deep semantic understanding. We note that our approach performs particularly well on issues requiring multi-line reasoning (and editing) and edge-case handling, suggesting that incorporating issue and code semantics into APR pipelines can lead to robust and semantically consistent repairs.
\end{abstract}


\section{Introduction}

In recent years, LLMs have shown remarkable progress in self-contained, short context software engineering tasks such as function completion \cite{chen2021evaluatinglargelanguagemodels}, test case generation \cite{10.1145/3643769, tufano2021unittestcasegeneration}, function summarization \cite{sun2024sourcecodesummarizationera}, and bug resolution \cite{10.1145/3611643.3613892}. Given this success, recent works such as SWE-Bench \cite{jimenez2024swebenchlanguagemodelsresolve} have introduced challenging software engineering benchmarks requiring LLMs to reason at larger repository-level contexts. In particular, these benchmarks introduce the task of \textit{repository-level program repair}. SWE-Bench curates such program repair tasks from resolved user-submitted issues in open-source repositories. Each repair task has three components: (i) a natural language user issue, (ii) a ground-truth developer patch, and (iii) test cases that reliably measure the efficacy of a patch. 

Repository-level program repair poses unique challenges for LLMs: (a) accurately interpreting the user-reported issue, (b) retrieving and reasoning over relevant code context—including dependencies across multiple functions and files, and (c) generating patches that meet the standards of expert developers by resolving the issue without breaking existing functionality. Advancing this task is essential for reducing developer burden and accelerating issue resolution. As a result, the task has attracted growing interest from both academia~\cite{zhang2024autocoderoverautonomousprogramimprovement, ruan2024specrovercodeintentextraction, xia2024agentlessdemystifyingllmbasedsoftware, wang2025openhandsopenplatformai, yang2024sweagentagentcomputerinterfacesenable} and industry~\cite{AWSAmazonQDeveloper, Cognition}, particularly through the development of agentic solutions.


Existing repository-level program repair methods fall into two main categories: open-process agent-based and workflow-based~\cite{anthropic}. \textit{Agent-based} methods treat LLMs as completely autonomous agents -- equipping them with tools to interact within a pre-defined environment.  While effective~\cite{wang2025openhandsopenplatformai, yang2024sweagentagentcomputerinterfacesenable, Cognition}, they often suffer from non-deterministic behavior, limited transparency, and poor controllability—making them difficult to trust, reproduce, or integrate into structured development workflows.


\textit{Workflow-based} approaches ~\cite{zhang2024autocoderoverautonomousprogramimprovement, ruan2024specrovercodeintentextraction, xia2024agentlessdemystifyingllmbasedsoftware}, on the other hand, propose guiding LLMs through pre-defined, task-specific procedures for repository-level program repair. These approaches leverage specialized tooling and exhibit greater determinism compared to open-ended, agent-based methods. While effective in addressing the primary issue, we observe that they often produce \textit{hyper-localized} patches — either 
(a) overfitting to specific bug-producing examples in the user's issue description, or (b) resolving the main issue but overlooking edge cases or broader consistency requirements. This limitation arises from two key challenges: (i) insufficient semantic understanding of the user issue and its associated code dependencies, and (ii) limited use of available execution semantics.


\textbf{Our work.}  To address the limitations of existing approaches, we propose \ToolName{}—a \textit{semantics-aware}, workflow-based technique that produces patches which are both \textit{complete} and \textit{consistent}. While retaining the reliability of workflow-based methods, \ToolName{} achieves superior performance by deeply understanding the user issue, code implementation, and execution behavior. \ToolName{} consists of two key components:

\begin{itemize}[leftmargin=*,noitemsep,topsep=0pt]
    \item \textbf{Bug Localization via Execution Semantics}: This component leverages a suite of methods—including issue reproduction, execution trace analysis, and spectrum-based fault localization—to identify buggy code regions relevant to the reported issue.
    
    \item \textbf{Patch Refinement via Semantic Analysis}: Given a candidate patch (often incomplete or inconsistent), this component refines it using:
    \begin{itemize}[leftmargin=*,noitemsep,topsep=0pt]
        \item \textbf{Issue Semantics}, which extracts intent and context from the user-reported issue.
        \item \textbf{Code Semantics}, which captures and reasons about underlying code dependencies affected by the patch.
    \end{itemize}
\end{itemize}

Together, these components enable \ToolName{} to generate high-quality fixes grounded in semantic understanding.


To this end, we make the following contributions. 
\begin{enumerate}[leftmargin=*,noitemsep,topsep=0pt]

\item \textit{\ToolName{}} - We introduce \ToolName{}, a novel pipeline that excels at creating complete patches that align with developer expectations. \ToolName{} pipeline consists of an improved bug localization agent with execution semantics, a novel patch repair agent (with issue and code semantics), and new patch reviewing and aggregator agents.

\item \textit{Issue Semantics and Program Semantics} - We highlight the importance for agents to develop a deeper understanding of code semantics (both issue and underlying program semantics). We validate the need for such semantics through extensive ablations.

\item \textit{Execution Semantics} - We showcase the utility of using execution semantics for the accurate localization of buggy context for the relevant issue. 

\item \textit{Evaluation.} \ToolName{} scores 44.66\% on SWE-Bench Lite, the highest among workflow-based methods. With full semantic integration in a multi-agent setup, it shows the potential of achieving around 51.33\%—leading the leaderboard across workflow-based and open-process agentic methods, with only a learning-based agent (a different category) ahead.
\end{enumerate}


We hope that future works incorporate elements of our approach into their pipeline by leveraging semantics-aware reasoning to resolve repository-level issues. In what follows, we discuss important related work (\S\ref{sec:related_work}), motivate our methodology with the help of an example (\S\ref{sec:motivation}), explain our specific methodology (\S\ref{sec:methodology}), and report our experimental results on SWE-Bench Lite (\S\ref{sec:experiments}). We also provide the prompts used in (\S\ref{sec:prompts_used}).




\begin{figure*}[t]
    \centering
    \includegraphics[width=\textwidth]{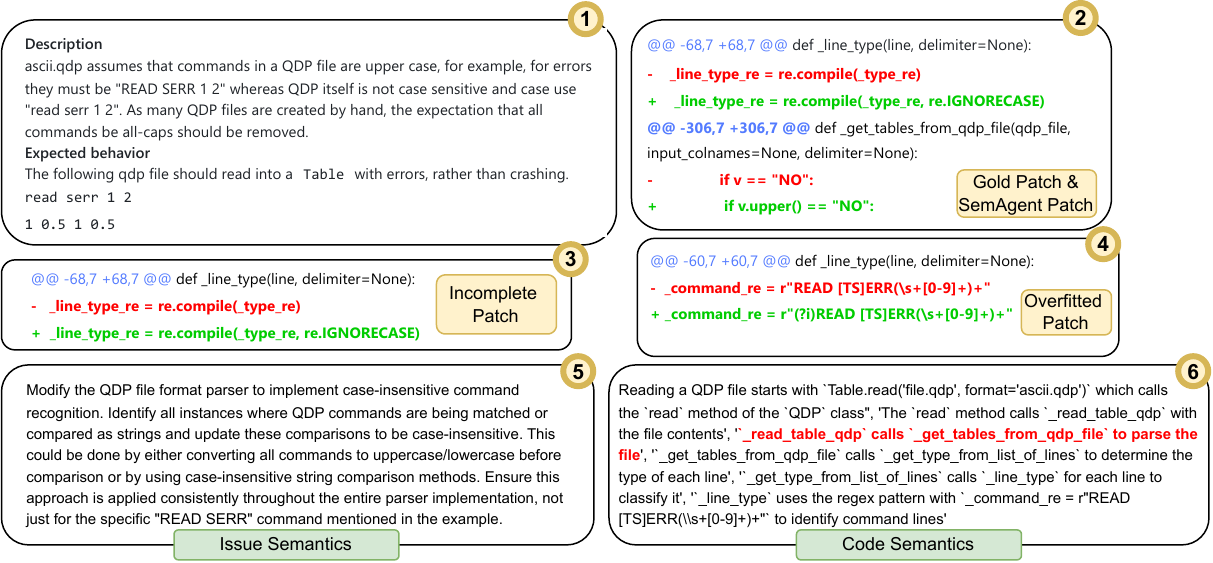}
    \caption[]{Motivating Example - \blackcircled{1} is the issue description, \blackcircled{2} is the developer patch \& SemAgent patch, \blackcircled{3} and \blackcircled{4} are incomplete/overfitted patches, \blackcircled{5} is the \ToolName{} generated issue semantics, \blackcircled{6} is the \ToolName{} generated code semantics.
    }
    \label{fig:motivating}
\end{figure*}

\section{Related Work}
\label{sec:related_work}

With the advent of LLMs, there has been a body of work on applying LLMs for downstream software engineering tasks, including code generation \cite{chen2021evaluatinglargelanguagemodels}, test creation \cite{kang2023largelanguagemodelsfewshot}, automated code review \cite{li2022automatingcodereviewactivities} and program repair \cite{gao2022programrepair}.
In particular, as LLMs have grown more capable and performant, there has been accute interest in applying LLMs as agents on tasks like repostiory-level program repair. Benchmarks like SWE-Bench and others \cite{jimenez2024swebenchlanguagemodelsresolve, rashid2025swepolybenchmultilanguagebenchmarkrepository, zan2025multiswebenchmultilingualbenchmarkissue, mathai2024kgym} have made it possible to conduct agent-related research by leveraging execution feedback through the automatic evaluation of test-cases. 
These benchmarks have consequently inspired extensive research into various techniques for generating patches that address reported issues. We categorize this related work into two primary approaches: (i) Open process agent-based frameworks and (ii) workflow-based frameworks.

\textbf{Agent-Based Frameworks.} 
The first category of approaches frames LLMs as agents equipped with tools to interact within a predefined environment~\cite{Cognition, wang2025openhandsopenplatformai, yang2024sweagentagentcomputerinterfacesenable}. These agents plan actions, execute them, and adapt based on environmental feedback. Tools often mimic developer activities such as searching, editing, and running code. For instance, SWE-Agent~\cite{yang2024sweagentagentcomputerinterfacesenable} provides LLM-friendly interfaces for navigation, editing, and execution, while OpenHands~\cite{wang2025openhandsopenplatformai} supports a broader action space using CodeAct~\cite{wang2024executablecodeactionselicit}, including Internet access. A defining feature of agent-based methods is their lack of a fixed procedure (i.e., open process)—actions are determined dynamically based on ongoing interaction and feedback.


\textbf{Workflow-based Frameworks.} 
The second category includes workflow-based methods that follow a pre-defined Search-Edit-Test sequence for resolving repository-level issues. These approaches first localize buggy code, then suggest edits, and finally validate them via regression or reproducer tests. Examples include Agentless ~\cite{xia2024agentlessdemystifyingllmbasedsoftware}, which guides agents through repository structure to identify context, propose edits, and run tests; AutoCodeRover ~\cite{zhang2024autocoderoverautonomousprogramimprovement}, which integrates program analysis tools for context lookup; SpecRover ~\cite{ruan2024specrovercodeintentextraction}, which adds specification generation and patch review, and Patch Pilot \cite{li2025patchpilot} which additionally includes patch refinement.



We build on workflow-based approaches as they provide better experimental control and reproducibility \cite{li2025patchpilot}. Beyond standard Search-Edit-Test steps, our pipeline introduces key semantics reasoning components: (a) execution-guided bug localization, (b) issue understanding, (c) semantic modeling of code using issue context, (d) fine-grained fix generation, and (e) minimal patch refinement via review.  
We motivate this design with an example bug from SWE Bench.


\section{Motivation}
\label{sec:motivation}


In this section, we (i) present a motivating example from SWE-Bench and highlight common patches from other SOTA agents (\S\ref{sec:astropy_issue}), and (ii) discuss the need for a deeper understanding of execution (\S\ref{sec:execution_semantics}), issues (\S\ref{sec:issue_semantics}), and code semantics (\S\ref{sec:code_semantics}).

\subsection{Astropy Issue}
\label{sec:astropy_issue}


To motivate our approach, we examine \texttt{astropy-14635}, a SWE-Bench Lite task that challenges many SOTA methods. As shown in Figure~\ref{fig:motivating}, the user-reported issue (box \blackcircled{1}) highlights a bug in the Astropy~\cite{astropy} library’s QDP file reader, which incorrectly assumes all input must be uppercase. The report includes a crashing example—``read serr 1 2''—and argues that the reader should accept lowercase input, as QDP is case-insensitive.  We observe that SOTA agents often overanalyze the issue description, producing patches that either are incomplete (i.e., fail to generalize to edge cases)  or overfit to the user inputs (Figure~\ref{fig:motivating}, boxes \blackcircled{3} and \blackcircled{4}, respectively). In contrast, \ToolName{} generates a consistent, generalizable patch aligned with the developer-written fix (box \blackcircled{2}) by leveraging execution, issue, and code semantics.


\subsection{Execution Semantics}
\label{sec:execution_semantics}

We leverage execution semantics including: (i) stack traces from issue reproducers, (ii) suspicious files identified in execution traces, and (iii) likely faulty code segments via spectrum-based fault localization (SBFL)~\cite{abreu2007accuracy}. This combination enables \ToolName{} to accurately localize the bug in our example to \texttt{qdp.py} and key functions such as ``\_line\_type", ``\_get\_lines\_from\_file", and ``\_get\_tables\_from\_qdp\_file'', aligning with the developer patch (box \blackcircled{2}). Execution semantics are especially valuable for complex repositories like Django,\footnote{\url{https://www.djangoproject.com/}} where issue reproduction requires configuring a full project environment.


\subsection{Issue Semantics}
\label{sec:issue_semantics}

The patch in box \blackcircled{4} illustrates \textit{overfitting}, where the agent fixes only the specific example input in the issue report while missing the broader underlying bug. We find this behavior common in SOTA agents, including Agentless~\cite{xia2024agentlessdemystifyingllmbasedsoftware}, which often hyper-localize based on the issue text. This leads to superficial fixes that act as band-aids rather than addressing root causes. We hypothesize that supplying agents with a deeper understanding of the issue (i.e., \textit{issue semantics}) can mitigate overfitting. Box \blackcircled{5} shows \ToolName{} generated issue semantics, which generalize the problem and explicitly guide edits across multiple components, enabling more robust and complete patches.

\subsection{Code Semantics}
\label{sec:code_semantics}

The patch in box \blackcircled{3} illustrates an \textit{incomplete} fix—while it addresses the core issue, it overlooks necessary updates elsewhere in the code (e.g., a related \texttt{if} condition). 
Advanced agents like SpecRover \cite{ruan2024specrovercodeintentextraction} are able to avoid ``overfitted'' patches and instead generate such ``incomplete'' patches. We hypothesize that fine-grained code semantics can help agents reason about these secondary changes. Box \blackcircled{6} shows a semantic trace from the QDP module; the red-highlighted step prompts \ToolName{} to identify and apply additional edits, resulting in a complete and consistent patch.

To this end, we show that \textit{execution}, \textit{issue}, and \textit{code} semantics are crucial for generating complete patches. 
We now describe \ToolName{}'s overall 
methodologies behind each component (\S\ref{sec:methodology}).

\begin{figure*}[pt]
    \centering
    \includegraphics[width=\textwidth]{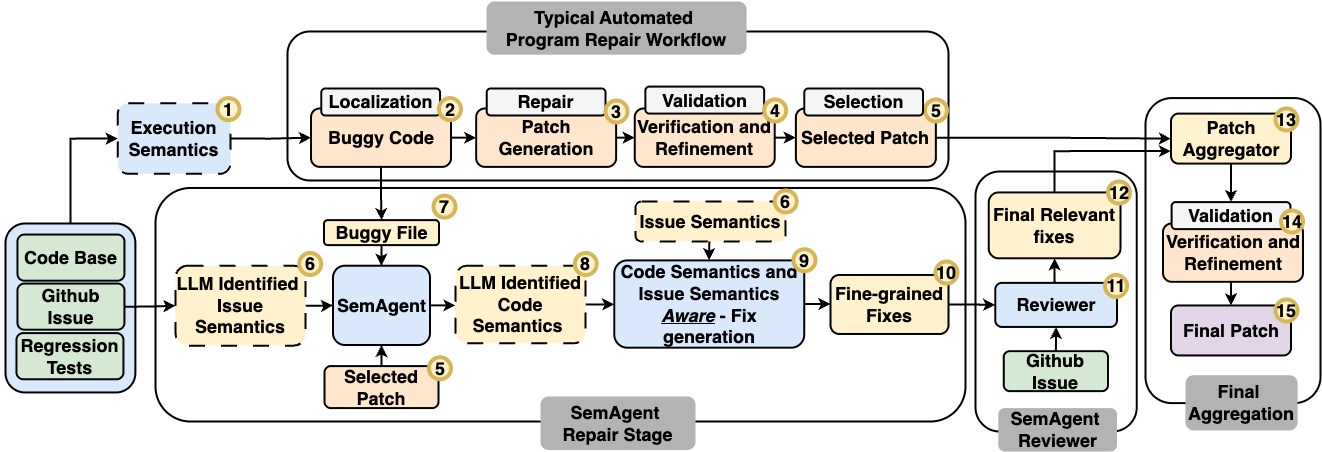}
    \caption[]{\textbf{Pipeline of \ToolName{}}. {\small It takes as input a \textit{codebase}, \textit{GitHub issue}, and \textit{regression tests}.
    First, it uses \textit{execution semantics} \circled{1} to localize the buggy code \circled{2} and apply a standard APR workflow—patch generation \circled{3}, validation \circled{4}, and selection \circled{5}—producing a candidate patch, which is often incomplete or inconsistent. Next, \ToolName{} refines this patch using issue \circled{6} and code semantics \circled{8}, extracted from the buggy file \circled{7} and candidate patch \circled{5}. 
    This yields fine-grained fixes \circled{10}, which a reviewer agent \circled{11} filters based on the issue, selecting relevant edits \circled{12}. These are aggregated with the candidate patch \circled{13} and refined using execution feedback \circled{14} to produce the final patch \circled{15}. All the semantic components are highlighted with dashed boxes.}}
    \label{fig:pipeline}
\end{figure*}

\section{Methodology}
\label{sec:methodology}


\textit{Problem Setup}: 
Given a Python codebase \(\mathcal{D}\) with regression tests \(\mathcal{T}\) and an issue statement \(\mathcal{I}\), the objective is to generate a patch $p$ such that the updated codebase \(\mathcal{D}'\) (after applying the patch $p$) passes an augmented test suite \(\mathcal{T}'\), which includes new or modified tests that reproduce the issue. Passing \(\mathcal{T}'\) confirms both issue resolution and preservation of original functionality.

We formally define \ToolName{} as a function \(\mathcal{F}(\mathcal{I}, \mathcal{D}, \mathcal{T}) \rightarrow p\), which generates the final patch $p$. First, \ToolName{} follows a standard Automated Program Repair (APR) pipeline--bug localization, patch generation, validation, and selection--to produce an initial patch \(p_{\text{ini}}\), which may be partial or incorrect. 
In this step, \ToolName{} provides semantics of code execution \((\mathcal{E})\) to facilitate this process.
To refine it, the repair module \(R\) takes \(p_{\text{ini}}\), the original codebase \(\mathcal{D}\), issue semantics \(\mathcal{I}'\) (computed via semantic analysis \(G(\mathcal{I}) \rightarrow \mathcal{I}'\)), and code semantics \(\mathcal{C}'\) (computed via semantic analysis \(S(\mathcal{C}) \rightarrow \mathcal{C}'\)), and generates a set of candidate patches \(\mathcal{P}' = \{p'_1, p'_2, \dots\}\).

These candidate patches are then passed to a reviewing agent \(\mathcal{V}\) that selects a subset \(\mathcal{P}''\) $\subseteq$ \(\mathcal{P}'\) of patches that are filtered based on relevancy in resolving the issue. From this subset, the final patch $p$ is constructed by combining the incomplete patch $p_{ini}$ and \(\mathcal{P}''\) and passing it through a patch generation agent that performs a final aggregation along with refinement and validation.

Our contributions are three-fold: (1) we design the repair stage, (2) formulate semantics of code, execution, and issues, and (3) implement a patch reviewer agent. We also show how these semantics enhance patch correctness, consistency, and context localization. Figure \ref{fig:pipeline} shows the complete outline of \ToolName{}.



\subsection{Execution Semantics for Bug Localization}

The execution semantics module (\circled{1} in Figure \ref{fig:pipeline}) localizes bugs by triangulating evidence from three sources: (1) crash reports from reproducers, providing coarse failure signals; (2) execution traces identifying actively involved files; and (3) Spectrum-Based Fault Localization~\cite{abreu2007accuracy}, which statistically ranks file suspiciousness using test outcomes. The localized regions are passed to an APR workflow (\circled{2} to \circled{5} in Figure \ref{fig:pipeline}) to generate an initial patch, which is then refined in \ToolName{}'s repair stage.

\subsection{\ToolName{}Repair Stage}
The goal of this stage is to refine an incomplete patch (\circled{5}) by generating a set of fine-grained fixes (\circled{10}) that resolve many incomplete patches and address the underlying issue. This stage comprises three components: (1) extracting issue semantics from the issue description (\circled{6}), (2) leveraging these to infer code semantics (\circled{8}), and (3) generating fixes using both (\circled{9}, \circled{10}).

\textbf{Generating Issue Semantics.} GitHub issues often include detailed descriptions, examples, and reproduction steps, but LLMs can overfit to specific artifacts (e.g., block \blackcircled{4} in Figure~\ref{fig:motivating}) or struggle with under-specified descriptions—both common in SWEBench Lite. To address this, we introduce a prompting step that guides the LLM to abstract beyond surface artifacts and produce structured semantic directions reflecting the core intent. These high-level signals ground the repair process, improving generalization and alignment with developer expectations.

\textbf{Generation of Code Semantics.} 
Understanding the issue alone is insufficient for generating robust patches—\ToolName{} must also capture how the code behaves in relation to the issue. This step builds a semantic model of the code’s intent, usage, and relevance to the bug.

\textit{Execution Flow Extraction.} We extract high-level execution flows relevant to the issue and candidate patch, capturing how control or data moves through the code. In our motivating example, four flows were generated (\blackcircled{6}) —e.g., reading a QDP table , processing QDP lines, writing QDP tables, and regex operations—each representing a distinct pathway for localized repair.

\textit{Step-Level Reasoning.} Each flow is decomposed into fine-grained, natural language steps (e.g., “read file line by line”, “extract header row”), which makes it easier to align reasoning with how developers typically conceptualize behavior. 
To map steps to code, we prompt an LLM with step descriptions and file content. The model returns a structured JSON mapping semantic roles to code snippets, enabling precise, context-aware localization.

\textbf{Semantics-Guided Patch Generation}: 
After localizing code snippets for each step, \ToolName{} generates candidate patches using three inputs: the localized code, step-level semantics, and issue semantics (\circled{6}). This ensures that fixes are grounded in both the high-level intent and functional behavior, avoiding superficial or overfitted changes. The resulting patches are semantically rich, context-aware, and aligned with developer expectations. They are then passed to the reviewer agent.

In the motivating example, for the red-highlighted step in \blackcircled{6}, \ToolName{} identifies the ``\_get\_tables\_from\_qdp\_file'' function and correctly adjusts the case-sensitive comparison \texttt{v == "NO"} to a case-insensitive form, resolving the issue as shown in \blackcircled{2}. To reduce costs at this token-intensive stage, \ToolName{} supports optional caching (Appendix~\ref{appendix:cache}), retrieving previously generated fixes for repeated issue–patch pairs and storing new ones for future reuse.

\subsection{Patch Reviewer Agent}
The final set of candidate patches is reviewed by an LLM-based reviewer agent (\circled{11}), which, given the issue description, determines patch relevance and provides justifications. Each patch is annotated with a boolean relevancy flag and an explanation, then passed to the final patch aggregation step. The reviewer agent uses a carefully designed prompt to align with the developer's intent and control the strictness of relevance assessment. This filtering step ensures only well-justified, semantically appropriate fixes are retained, increasing the likelihood of consistent and developer-aligned repairs.



\subsection{Patch Aggregator}
The selected patches (\circled{12}) are passed to a Patch Aggregator (\circled{13}), which deduplicates and merges overlapping edits to ensure unique, non-conflicting fixes. These are combined with the initial patch (\circled{5}) and refined via an execution loop to produce the final patch (\circled{15}).

Such an aggregator can be extended to a multi-agent setup where specialized agents generate complementary patches based on different kinds of semantics (see ablation: RQ2). The Patch Aggregator consolidates these diverse fixes into a coherent, validated patch that aligns with both the issue intent and code functionality.

\section{Experimental Setup}
\label{sec:experiments}
We address the following research questions:
\setlist{nolistsep}
\begin{itemize} [noitemsep,leftmargin=*]
\item \textbf{RQ1}: What is the impact of \ToolName{} on issue resolution?
\item \textbf{RQ2}: How useful are the individual components of \ToolName{} in patch generation?
\item \textbf{RQ3}: To what extent does incorporating Execution Semantics improve issue localization?
\end{itemize}

\subsection{Evaluation Benchmark}
We evaluate \ToolName{} on SWEBench Lite~\cite{jimenez2024swebenchlanguagemodelsresolve}, a benchmark of 300 real-world GitHub issues across 11 Python repositories. For each issue, \ToolName{} receives the issue description and pre-fix codebase with regression tests, and outputs a single patch in git diff format, which is evaluated for successful resolution.


\subsection{Setup}
\ToolName{} takes as input the issue description, codebase, and preloaded Amazon Q Developer reproduction tests~\cite{AWSAmazonQDeveloper}, following the workflow in Figure~\ref{fig:pipeline} to generate one patch per issue for evaluation. \ToolName{} adheres strictly to SWE-bench guidelines: it performs a single pass@1 attempt, uses no test-specific metadata or hints, avoids web access, and ensures all patches are self-contained and executable within the provided context.



\subsection{Baseline}

To evaluate effectiveness, we compare \ToolName{} against SpecRover\cite{zhang2024autocoderoverautonomousprogramimprovement}, which forms our architectural baseline and reports 37\% accuracy on SWEBench Lite\footnote{\url{https://www.autocoderover.net/}}. 
In addition to this we also compare against the best performing (as per SWEBench leaderboard) workflow based methods, Agentless \cite{xia2024agentlessdemystifyingllmbasedsoftware} and Agentic system OpenHands \cite{wang2024openhands}. 
We also include comparisons with other state-of-the-art APR methods, such as the agentic - CodeStory Aide, DARS Agent \cite{aggarwal2025dars}, and Globant Code Fixer Agent. These systems represent a range of workflow-based and agentic-based approaches, as well as a mix of open and closed-source implementations. This comprehensive comparison highlights the effectiveness and versatility of our method across diverse repair paradigms.

\subsection{Parameters}
\ToolName{} builds on workflow-based APR systems like SpecRover~\cite{ruan2024specrovercodeintentextraction}, reusing components for reproducer generation, context localization, and patching. We allow up to 3 pipeline retries, 15 localization rounds, and 10 reproducer attempts per issue, using Claude Sonnet 3.7 for its strong reasoning and code generation. To ensure stability, we fix the decoding temperature to zero and incorporate spectrum-based fault localization, improving reproducibility and reducing LLM-induced variance.



\section{Results}
\label{sec:results}

\subsection{RQ1: Impact on issue resolution}


Table \ref{tab:results_of_models_on_swebench} shows the performance of various approaches on the SWEBench Lite benchmark. Our method, \ToolName{}, successfully resolves \textbf{134} issues, with a resolution rate of \textbf{44.66\%}. This represents a substantial improvement over the SpecRover baseline, outperforming it by \textbf{23} additional resolved issues with a \textbf{7.66} percentage point increase in absolute resolution rate showing the effectiveness of incorporating issue and code semantics the program repair pipeline to improve performance.

\ToolName{} is the top purely workflow-based method and ranks second among all the open-source approaches on the SWEBench Lite leaderboard, behind only the agentic, fine-tuned DARS-Agent. Despite using only out-of-the-box models, \ToolName{} ranks 7th overall, highlighting the strength of incorporating code semantics in workflow methods. All results are from a single run.

\begin{figure}
\centering
    \includegraphics[scale=0.6]{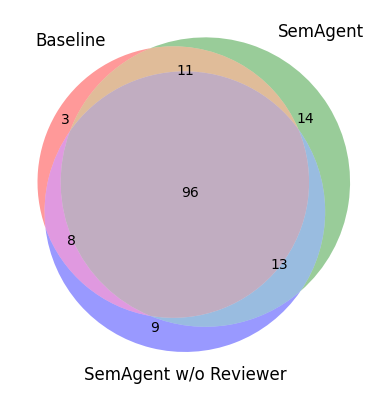}
    \caption[]{Overlap of Ablations - 154 issues resolved in total.}
    \label{fig:overlap}
\end{figure}

\ray{check}
Figure~\ref{fig:overlap} shows that our different design choices altogether can solve up to 154 issues. Thus, while extending our aggregator to a multi-agent (~\MToolName{}) setup, we could potentially solve up to 154 issues. 
The process can be further improved by smart sampling strategies, like Monte Carlo Tree sampling \cite{zhou2024languageagenttreesearch}.

\textbf{Time and Costs}: Our approach takes on average 16 minutes to solve an issue and costs 6.9\$ dollars per issue when run on the parameters mentioned in the Setup section. These costs can be reduced to 4.77\$ an issue across multiple runs when using issue semantics and repair stage caching. We also note that the median cost per issue is 4.87\$ which is significantly lower then the average due to a few costly outliers such as scikit-14087 (>30\$) that skew the average cost.

\begin{table*}[t]
  \centering
  \resizebox{.9\textwidth}{!}{
  \begin{tabular}{l | c | c | c}
    \hline
    \textbf{Category} & \textbf{Approach} & \textbf{Open-Source} & \textbf{SWEBench Lite Performance} \\
    \hline
    Workflow-based & AutoCodeRover & Yes & 30.67 (92) \\
    Workflow-based & SpecRover & Yes & 37\% (111) \\
    Workflow-based & Agentless & Yes & 40.67\% (122) \\
    Agentic & Openhands & Yes & 41.67\% (125) \\
    Agentic & CodeStory Aide & No & 43\% (129) \\
    \rowcolor{lightmauve} \textbf{Workflow-based} & \textbf{\ToolName{}} & \textbf{Yes} & \textbf{44.66\% (134)} \\
    Agentic & DARS Agent & Yes & 47\% (141) \\
    Agentic & Globant Code Fixer Agent & No & 48.33 \% (145) \\
    \rowcolor{lightmauve} \textbf{Workflow-based} & \textbf{\MToolName{}} & \textbf{Yes} & \textbf{51.33\% (154)} \\
  \hline
  \end{tabular}
  }
  \caption{RQ1.~Efficacy of approaches on SWEBench Lite}
  \label{tab:results_of_models_on_swebench}
\end{table*}

\subsection{RQ2: Ablation for patch generation}

\ToolName{} consists of two main components: the Repair Stage (with Issue and Code Semantics) and the Reviewer Agent. We ablate four variants to assess their impact: (1) baseline with no components, (2) only the Repair Stage with both semantics, (3) Repair Stage with only Code Semantics (using the raw issue description), and (4) the full system with both semantics and the Reviewer Agent. Comparing (1) vs. (2) shows the value of semantics; (2) vs. (3) isolates the impact of issue semantics; (2) vs. (4) highlights the reviewer’s contribution. Execution semantics ablations are deferred to RQ3, as they target localization, not patch generation.


\begin{table*}[ht]
\parbox{.48\linewidth}{
\centering
\resizebox{\linewidth}{!} {
\begin{tabular}{lc}
    \hline
    \textbf{Approach} & \textbf{Solve Rate} \\
    \hline
    Baseline & 37\% (111)\\
    w/ Repair Stage & 42\% (126) \\
    \rowcolor{lightmauve} \textbf{w/ Repair Stage + Reviewer} & \textbf{44.66\% (134)} \\
    \hline
  \end{tabular} 
  }
  \caption{RQ2.~Impact of Components on \ToolName{}Efficacy on SWEBench Lite} 
  \label{tab:results_of_componenets}
}
\hfill
\parbox{0.48\linewidth}{
\centering
\begin{tabular}{lc}
    \toprule
    \textbf{Approach} & \textbf{Solve Rate} \\
    \midrule
    w/o Issue Semantics & 50\% (25) \\
    \rowcolor{lightmauve} \textbf{w/ Issue Semantics} & \textbf{60\% (30)} \\
    \bottomrule
  \end{tabular}
  \caption{RQ2.~Impact of Issue Semantics on \ToolName{}Efficacy on a Random Set of 50 SWEBench Issues} 
  \label{tab:results_of_issue_semantics}
  }
\end{table*}


Table \ref{tab:results_of_componenets} presents results from ablation (1) and ablation (3) with the mentioned Repair Stage using both types of semantics. Ablation (2), shown in Table 4, is run on a smaller subset of 50 issues and demonstrates the effect of using issue semantics vs just the issue~\ray{can we get baseline numbers on Table 3 50 examples?}.

\textit{\textbf{Results}}: Table \ref{tab:results_of_componenets} shows the initial improvement on the baseline when incorporating semantics increasing performance by 5\%. Further analysis of the newly resolved issues shows that \ToolName{} added complete fixes that complimented the patch in half of these issues leading to a successful resolution and in the other half of issues, the results provided by \ToolName{} solve the complete issue and during the final patch generation phase the suggestions from \ToolName{} lead to a complete resolution of the issue. However upon closer inspection of some of the failed patches, we see that \ToolName{}adds extra fixes that can potentially break existing functionality or go above and beyond a simple fix required to solve the issues. Such fixes are now filtered by the reviewer agent leading to a 2.66\% increase in performance giving us our best performing approach. These results validate the importance of the components in SemAgent's workflow.

Table \ref{tab:results_of_issue_semantics} evaluates the impact of incorporating issue semantics in the Repair Stage of \ToolName{}, using 49 randomly selected low-cost issues along with the motivating example. The results reveal that 5 issues remained unresolved when issue semantics were excluded. A closer analysis indicates that, in these cases, the LLM fails to produce meaningful modifications beyond the initially generated incorrect patch. Notably, in the motivating example, the absence of semantic context leads to inadequate handling of consistency and edge cases showing the critical role issue semantics play in \ToolName{}’s pipeline.


\subsection{RQ3: Impact on Issue localization.}

Accurate localization is critical in workflow-based repair, as early errors propagate downstream. Incorporating execution semantics improves localization by 3.67\% (Table~\ref{tab:execution_impact}). 

\begin{table}[ht]
  \centering
  \begin{tabular}{lc}
    \hline
    \textbf{Approach} & \textbf{Localization \%} \\
    \hline
    w/o Execution Semantics & 82\% \\
    \rowcolor{lightmauve} \textbf{w/ Execution Semantics} & \textbf{85.67\%} \\
    \hline
  \end{tabular}
  \caption{RQ3.~Impact of Execution Semantics on \ToolName{} Issue Localization Accuracy}
  \label{tab:execution_impact}
\end{table}


We observe that for repositories like Django with complex setup requirements, spectrum-based fault localization 
are particularly effective. However, for projects with simpler setups such as Matplotlib and scikit-learn, stack traces and suspicious files derived from execution traces tend to yield more accurate localization results along with richer contextual information to aide localization. Further analysis also reveals that execution semantics also enhance determinism, enabling better caching, reducing average resolution time (17 to 16 minutes), and improving stability—making them a key component of the \ToolName{} workflow.

\section{Conclusion}

In this paper, we present \ToolName{}, a workflow based APR workflow that achieves 44.66\% accuracy on SWE-Bench Lite leveraging issue, code, and execution semantics to generate complete and consistent patches for repository-level repair, and highlight the importance for agents to incorporate semantics into their workflow to develop a deeper understanding of both issue and code semantics to improve performance.
\newpage
\section{Limitations}

While \ToolName{} demonstrates promising results, several limitations remain. First, the current implementation is restricted to single-file fixes, which limits its applicability to issues that span multiple files or directories and also affects context localization during step-level repair. In its current form, the system may miss relevant code outside the immediate scope of the identified file. One possible direction is to integrate a repository-level map that can be used to - (1) identify call chains from code subspaces through different repositories and files, and (2) retrieve context relevant to each step from code subspaces. Scaling the approach to operate effectively across these larger scopes remains an open challenge and we hope to fix this in future works.

This system is also expensive to run, with an average cost of approximately \$6.9 per issue when using Claude Sonnet 3.7 at a cost of \$3 per one million tokens, although this cost significantly decreases with multiple runs due to \ToolName{}'s repair stage caching system or when using cheaper more affordable models we acknowledge that costs can further be reduced using efficient prompt caching and optimizing intermediate steps to reduce the number of LLM calls and tokens used.

\section{Potential Risks}

\ToolName{} shares common risks associated with any automated program repair system deployed in real-world scenarios. These include the injection of vulnerabilities i.e code that introduces new security flaws in the codebase, failure to account for edge cases, and the possibility of unintended consequences in production environments. While our approach tries to minimize errors such as unaccounted edge cases, it remains crucial that the output is thoroughly reviewed before integration without blind overtrust in Autonomy.


\bibliography{custom}

\appendix

\section{Appendix}
\label{sec:appendix}

\subsection{Modifications To The Reproducer}
\label{appendix:reproducer}

SpecRover contains a Reproducer Agent that takes in the issue description and generates a standalone python file that when executed in the repository reproduces the issue. The generated reproducer is determined to be successful if it raises an Assertion Error and exits with a non zero code with the prompt generating the reproducer specifically mentioning these conditions. In our experiments we have noticed that this system exit sys.exit() usually in a caught error can return stack traces that aren't helpful in localizing the faulty issues. We first improve the feedback prompt to specifically mention if a reproducers stack trace doesn't have an assertion error or exits with a zero code which significantly speeds up how many rounds it takes to generate a successful reproducer.

Next we add another component that takes in the best possible reproducer and runs it through an execution feedback loop with the goal of refining the reproducer and modifying it generate a more useful stack trace that can better help with issue localization. The feedback loop consists of passing the reproducer, the stack trace and the issue to the LLM and prompt it using a custom prompt that focuses on refining the reproducer to - (1) better reproduce the issue and (2) to remove potential try catch, system exit codes and assertion errors and generate a reproducer that gives the most detailed stack trace errors and incorporate an LLM call that decides if the stack trace is sufficiently useful or not. 

We find that this step significantly improves the quality of stack trace errors produced and constitutes the first step of our execution semantics.

\subsection{Identification of Suspicious Files From the Execution Trace}
\label{appendix:trace}

Once we have this refined reproducer, we run it and process the execution trace to identify all the relevant files that were called in the reverse order of contact i.e, the first file in the trace is the file that was last used before the crash occurred. We then identify all the files in the repository and store them in a set. Next we filter out the files obtained from the execution trace an only include those files that are both in the execution trace as well as in the repository, this gives us an ordered list of suspicious files that were last used when the crash happened. 

We then take the top 7 files and give this to the localization agent along with a prompt explaining why they are relevant. This is the second step in the execution semantics. The final step is running Spectrum-based Fault Localization and obtaining the top 5 most suspicious functions along with their class and file and sending this to the localization agent with a suitable prompt that similarly explains why these files are useful and how they were obtained.

\subsection{Repair Stage Caching}
\label{appendix:cache}

When the repair stage is executed for a given issue (\textit{I}) on a buggy file (\textit{F}) using an initial patch (\textit{P}), it produces a set of candidate patches (\textit{$p_i$}). We store these results in a cache, using the tuple (\textit{I}, \textit{F}, \textit{P}) as the key and the resulting patches (\textit{$p_i$}) as the value. This caching mechanism helps avoid redundant repair runs for previously encountered inputs, leading to significant cost savings. The cache is intended to be used primarily when running powerful models with temperature set to 0, to minimize variance. Use of the cache is \textbf{entirely optional} and was solely designed to help reduce cost.

\subsection{Prompts Used:}
\label{sec:prompts_used}

\begin{tcolorbox}
    [title={Generic System Prompt for Issue Semantics, Workflow generation, Context Retrieval and Fix Generation}]
You are a software developer working on fixing an issue in your code base. Another software developer has looked into this issue and has given 
a fix that solves the main issue or solves it to the best of the developers ability. Now, your goal is to analyze the possible call chains in a flle, go through and analyze them 
and determine if any additional changes need to be made in addition to the main change to maintain consistency in the code base while 
also fixing any edge cases or existing functionality that might have broken.
\end{tcolorbox}

\begin{tcolorbox}
[title={Prompt To Generate Issue Semantics}]
The issue might show a very specific example failing or not clearly explain the intent of the developer,  your goal is to understand the exact intent of what the issue conveys, and explain if this issue is part of a broader issue and if it is generalizable, and explain what the code should reflect to solve this issue as a whole,  then summarize everything you've said. Finally end it by by giving a paragraph that would direct an AI agent to fix changes that need to be made in the code base,  be as general as possible. The given issue you need to generalize is: "\{\textcolor{blue}{issue statement}\}". The final paragraph that directs an AI agent  should be in a <directions> ... </directions> block.
\end{tcolorbox}

\begin{tcolorbox}
[title={Prompt To Generate Workflows}]
    You are an AI assistant trying to solve an issue: \{\textcolor{blue}{issue semantics}\}, you have correctly figured out the main buggy file and fixed it using the 
        patch: "\{\textcolor{blue}{initial patch}\}" you however miss other changes required to maintain the consistency of the fix throughout the file. the content of the 
        file: "\{\textcolor{blue}{buggy file name}\}" is: "\{\textcolor{blue}{code in buggy file}\}". Your task is to figure out the different flows of execution of the file along with the names of the methods called.
        We will then look at each execution flow and determine if there needs to be other fixes in addition to the main fix to maintain consistency.\\
\\
        Each flow must be in the following format: \\ \\
        <flow> \\
        <step> describe the approach...</step> \\
        <step> ... </step> \\
        ... \\
        </flow> \\

        <flow> \\
        <step>...</step> \\
        <step>...</step> \\
        ... \\
        </flow> \\
\\
        ...\\
\\
        Each flow much be encased in <flow> </flow> and each step must be in <step> </step>. \\
\end{tcolorbox}

\begin{tcolorbox}
    [title = {Prompt To Get Relevant Context For A Step}]
    You are analyzing a call chain, and are given steps that describe steps in a flow such as `\_read\_table\_qdp()` calls `\_get\_tables\_from\_qdp\_file()`,
        along with the entire code in the file that the call chain resides in. Your goal is to understand the relevant code snippets that need to be 
        extracted in order to analyze this step and solve an issue. Your output should be a JSON with the key describing the code snippet, and the 
        value being the code in the code snippet. NOTE: If you are giving the code of a method then you must give the ENTIRE code in the method. Also note that the key and values of the JSON must be strings.
        For example, for the `\_read\_table\_qdp()` calls `\_get\_tables\_from\_qdp\_file()` step, the key should be something like "code in the \_get\_tables\_from\_qdp\_file() method",
        and the value should be the code in the \_get\_tables\_from\_qdp\_file method.
        the resulting JSON structure should look like:
        {{
            "code in the \_get\_tables\_from\_qdp\_file() method": Get all tables from a QDP file.  Parameters .... (rest of the \_get\_tables\_from\_qdp\_file method code)
        }}
        The file\_content is: "\{\textcolor{blue}{File Content}\}", and the step in the call chain is: "\{\textcolor{blue}{Step}\}".
        Your goal is to find the relevant code snippets as mentioned above.
        NOTE: Try to focus the majority at a method level of granularity, however you can also use a class level or expression level of granularity.
        This information will be given to another AI agent who will suggest code changes to fix the bug and maintain consistency throughout the file.
\end{tcolorbox}

\begin{tcolorbox}
[title={Prompt To Generate Fixes}]
    You are given an execution 
            flow: <flow>"\{\textcolor{blue}{Flow} \}"</flow> and general directions on how to fix it: <directions>" {Issue Semantics} "</directions>. In this flow we focus on the step: <step>""\{\textcolor{blue}{Step}\}"</step>
            .the relevant code snippets are: <codesnippets>""\{\textcolor{blue}{Relevant Code Snippets}\}"</codesnippets>, and another AI agent has given an initial patch: <patch>""\{\textcolor{blue}{Initial Patch}\}"</patch> that could potentially have solved the bulk of the issue. Your goal is to analyze these code snippets and identify changes that need to be made to maintain consistency 
            with the general directions provided to solve the overall issue. 
            These can be minor or major changes that try to maintain consistency in the file, 
            fix edge cases that might have been missed, 
            or fix modified code that might break important existing functionality. 
            Please be EXTREMELY thorough, go through each line of context given to make 
            sure that you have not missed anything that needs to be changed. You can give your reasoning, however your final changes must be in 
            <changes>...</changes> angle brackets, with the original code, patched code and reasoning, preferably in a <original>...</original> 
            <patched>...</patched> <reason>...</reason> format. If there are no changes to be made then the output must be "No changes" in the angle brackets, 
            i.e <changes>No changes</changes>.
            
\end{tcolorbox}

\newpage

\begin{tcolorbox}[breakable, title = {Prompt For Repair Stage Fixes Reviewer}]
\textbf{System Prompt:}\\ \\
You are an expert software code reviewer who evaluates suggestions made by multiple software engineers attempting to solve a specific issue in a code file.\\

Your role is to:\\
- Understand the issue in depth.\\
- Analyze each suggestion carefully, identifying which are necessary, which are helpful, and which may be incorrect or unnecessary.\\
- Apply clear and well-reasoned judgment to select the best combination of suggestions to fully and correctly resolve the issue.\\

You prioritize correctness and consistency in your review. Use precise reasoning when justifying your decisions.\\

\textbf{User Prompt:}\\\\
You are reviewing and trying to solve the following issue in a code file:

        <issue> "\{\textcolor{blue}{Issue Statement}\}" </issue>\\

        The full content of the file is:\\

        <file content> "\{\textcolor{blue}{File Content}\}" </file content>\\

        A number of software engineers have provided suggestions, each aiming to solve the issue along with a starting fix to build on that most likely fixes the bulk of the issue:\\

        <starting fix> "\{\textcolor{blue}{Patch Content}\}" \\</starting fix>\\

        These suggestions consist of the original code, the patched code, and the reasoning for why this change was suggested.
        Your goal is to filter out unnecessary suggestions while keeping the useful ones.\\

        Useful suggestions fall into one of these categories:\\
        - Suggestions ensuring the consistency of the fix throughout the file.\\
        - Suggestions identifying edge cases that might have been missed by the starting fix.\\
        - Suggestions fixing code that may have been broken by the starting fix.\\
        - Suggestions that solve the core issue if the starting patch identifies a wrong fix.\\

        where as,\\
        
        Unnecessary suggestions fall into one of these categories:
        - Changes that break existing functionality. Use your best judgment when deciding this.\\
        - The addition of unnecessary try catch statements or assertions which look good in theory but are unnecessary as they might break unit testing for that particular functionality and could have already been caught else where.\\
        - If the issue is incredibly simple to fix, then avoid overtly complex suggestions which are prone to break some existing functionality.\\

        You will be given "\{\textcolor{blue}{Number of Patches}\}" suggestions in the format of:\\
        
        0: <original> ... </original> <patched> ... </patched> <reason> ... </reason>\\
        1: <original> ... </original> <patched> ... </patched> <reason> ... </reason>\\ \\
        and so on, with 0: representing the 0th patch, 1: the 1st, etc...\\

        Your output must be a JSON object in the following format:
\\
        \{\{ \\
            "0": \{\{ \\
                "reason": "Explanation of why this suggestion is necessary or not.", \\
                "required": "Required or Not Required", \\
            \}\}, \\
            "1": \{\{ \\
                "reason": "Explanation of why this suggestion is necessary or not.", \\
                "required": "Required or Not Required", \\
            \}\}, \\
            ... \\
        \}\} \\
\\
        - Each key corresponds to the suggestion ID. \\
        - The "reason" field must contain a concise, clear justification for why the suggestion is necessary or not. \\
        - The "required" field should be `Required` if the suggestion is required to solve the issue, or `Not Required` if it is not needed. \\
        - Always begin with your reasoning in the "reason" field, followed by the decision in the "required" field. \\

        The suggestions are: \\

        "\{\textcolor{blue}{Patches Enumerated One By One}\}" \\

        If there are no suggestions then return None.
        """

\end{tcolorbox}

\end{document}